**Shear and longitudinal viscosity of non-ionic $C_{12}E_8$ aqueous solutions**

G. D'Arrigo*
*INFM, Dipartimento di Energetica, Università di Roma “La Sapienza", Via A. Scarpa 16, 00161 Roma, Italy*

G. Briganti
*CRS-SOFT, Dipartimento di Fisica, Università di Roma “La Sapienza", Piazzale A. Moro 2, 00185 Roma, Italy*

M.Maccarini
*Current Address: Institut für Angewandte Physikalische Chemie, Ruprecht Karls Universität Heidelberg, Im Neuenheimer Feld 253, D-69120 Heidelberg, Germany*

## ABSTRACT

We present an extensive set of measurements of steady shear viscosity ($\eta°_s$), longitudinal elastic modulus (M') and ultrasonic absorption ($\alpha$) in the one-phase isotropic liquid region of the nonionic surfactant $C_{12}E_8$ aqueous solutions. Within a given temperature interval, this phase extends along the entire surfactant concentration range, that could be fully covered in the experiments. In agreement with previous studies, the overall results support the presence of two separated intervals of concentration corresponding to different structural properties. In the surfactant-rich region the temperature dependence of $\eta°_s$ follows an equation characteristic of glass-like systems. The ultrasonic absorption spectra show unambiguous evidence of viscoelastic behaviour that can be described by a Cole-Cole relaxation formula. In this region, when both the absorption and the frequency are scaled by the static shear viscosity ($\eta°_s$), the scaled attenuation reduces to a single universal curve for all temperatures and concentrations. In the water-rich region the behaviour of $\eta°_s$, M' and $\alpha$ are more complex and reflect the presence of dispersed aggregates whose size increases with temperature and concentration. At these concentrations the ultrasonic spectra are characterized by a multiple decay rate. The high frequency tail falls in the same frequency range seen at high surfactant content and exhibits similar behaviors. This contribution is ascribed to the mixture of hydrophilic terminations and water present at the micellar interfaces that resembles the condition of a concentrated polymer solution. An additional low frequency contribution is also observed, which is ascribed to the exchange of water molecules and/or surfactant monomers between the aggregates and the bulk solvent region.

* Author to whom correspondence should be addressed: Giovanni D'Arrigo, Dipartimento di Energetica, Università "La Sapienza", Via Scarpa 16, 00161 Roma, Italy. FAX 39-6-44240183. Electronic address:giovanni.darrigo@ uniroma1.it

## 1. INTRODUCTION

The properties of aqueous solutions of non-ionic surfactants of the oligooxyethylene glycol have been largely studied in the last years. These surfactants are commonly denoted as $C_iE_j$, where $C_i$ ($\equiv C_iH_{2i+1}$) indicate the linear hydrocarbon chain (hydrophobic tail) and $E_j$ [$\equiv (OCH_2CH_2)_jOH$] the hydrophilic head composed by j oxyethylene (EO) units.

Unlike simple binary mixtures, on varying temperature and concentration $C_iE_j$ solutions display a rich variety of thermodynamic phases[1] such as anisotropic phases (lamellar, cubic, hexagonal, solid), bimodal curves and critical points, as well as isotropic phases. Experimental and theoretical efforts have been widely carried out on this non-ionic surfactant family, but most of the works concerns the dilute and semidilute regime (up to about 20 wt%: surfactant mass fraction), where the isotropic phase can still be described by a continuous dispersion of micelles. In this region a considerable amounts of reports demonstrated a significant micellar growth with temperature and concentration (see, for example, ref.[2,3,4,5,6]).

However the evolution toward the pure surfactant liquid state as the water content reduces is not yet clear. Some experimental studies using various scattering techniques (X-rays and neutrons[7], Raman[8] and light[9]) showed that these solutions exhibit at least two characteristic concentration ranges, below and above an amphiphile volume fraction $\phi^*$ that depends on the hydrophilic/hydrophobic volume ratio. Below $\phi^*$ the micellar structure is retained while, at volume fractions higher than $\phi^*$, it is envisaged an evolution toward a block-polymer melt. Apart from the previous scattering experiments, few other studies were performed in the entire concentration range[10].

In this work we report extensive measurements of shear viscosity, ultrasonic velocity and attenuation on the whole concentration range of $C_{12}E_8$ water solutions. The volumetric properties, needed in our analysis, have been published elsewhere[11]. In a previous work[5] we studied $C_{12}E_6$ aqueous solutions using the same techniques but, for this surfactant, the concentrated region was experimentally inaccessible due to the presence of hexagonal and lamellar anisotropic phases. On the contrary, $C_{12}E_8$ water solutions exhibit an isotropic phase that extends along the entire surfactant concentration range between $T_M \approx 59°C$, the melting point of the liquid crystalline hexagonal phase $H_1$, and $T_{crit} \approx 80°C$, the critical point temperature (see the phase diagram in Figure 1).

The aim of the article is to probe the structural conditions of $C_{12}E_8$ solutions, combining the results of static (density, shear viscosity and longitudinal moduli) and dynamic (sound attenuation) properties. The frequency window of the ultrasonic technique generally covers the range from about 1 MHz to 1 GHz, thus lying between that of mechanical techniques ($0 \div 10^5$Hz) and those related to molecular properties typical of NMR, Brillouin, neutron, Raman and x-ray scattering ($10^8$ to $10^{14}$ Hz).

The article is organized as follows. In section 2 the experimental methods and the sample preparations are described. The results are reported in Section 3, divided in three subsections: the first two are devoted to the analysis of the viscosity and longitudinal elastic moduli, the third to the sound attenuation. This subsection contains a brief presentation of the basic equations concerning the attenuation, its relation with the viscosity, and the definition of a specific observable used in our analysis: the ratio between the experimental attenuation and the Navier-Stokes limit. In section 4 the discussion and conclusion are presented.

## 2. EXPERIMENTS

The $C_{12}E_8$ non-ionic amphiphile used in this experiment (purity ≈ 99%, M.W. 490) was purchased by Nikko Chemical, Japan and was used without further purification. Bidistilled deionized gas-free water was used and the samples were prepared by weight just before the beginning of each measurement.

In micellar aqueous solutions and other supramolecular systems it is customary to plot properties as a function of volume fraction ($\phi$) of aggregates. This choice is suitable to test theoretical models. However, evaluation of $\phi$ from known amounts of components at the mixing (e.g. volumes or masses) is not feasible since mixing is not ideal. In micellar systems, for example, the volume fraction of aggregates does not correspond to the volume of surfactant molecules because of hydration. A good approximation is given by the relation $\phi=1-(1-wt\%)\rho/\rho_w$, where $\rho$ and $\rho_w$ are, respectively, the density of solution and water solvent, and wt% is the surfactant mass fraction. Since in our solutions it is $\rho \approx \rho_w$ (within 2%), we approximate $\phi \approx wt\%$.

Kinematic shear viscosity ($\nu$) was measured with standard calibrated Ubbelhode tubes. Using viscosimeters with different constants (100-500 $s^{-1}$) we found nearly equivalent values indicating that in this range the measured viscosity represents the static (zero frequency) values. The steady dynamic shear viscosity ($\eta_s^\circ$) was obtained by the kinematic values multiplying by measured densities ($\rho$) reported elsewhere[11]. Viscosity was measured in the samples of concentration 2.5, 9, 20, 30, 65, 80, 90, 100 wt% in the temperature interval 15-70°C. Measurements at 40 and 50 wt% were not considered in the viscosity analysis since they were affected by large uncertainties, likely due to the closeness to the hexagonal phase. Nevertheless the viscosities at these concentrations were interpolated and used in the sound attenuation analysis. Additional viscosity data at 1 and 5 wt%, useful for the analysis, were taken from the literature[12].

Ultrasonic measurements in the frequency range 5-200 MHz were performed by a standard pulse technique employing a Matec equipment and a variable path ultrasonic cell with two 5 MHz fundamental quartz transducers. Details of the experimental procedures to get sound velocity and attenuation coefficient are reported elsewhere[5]. Sound velocity (c) was measured by an overlapping pulses method that assures a detection of velocity dispersion within ±1m/s in the range 5-75 MHz. However, due to highly attenuating samples, dispersion was measured, as a rule, in the 5-45 MHz range. The accuracy of the absolute velocity (at fixed frequency) was estimated to be ±0.2 m/s. Accuracy in the measured sound attenuation coefficient ($\alpha$) depends on $\alpha$ itself. At the lowest experimental frequency (5MHz) it was estimated to vary between 5% (highly attenuating samples) to ~10% (low absorbent samples). Better accuracies are obtained at higher frequencies. The samples investigated in the ultrasonic experiments were 9, 20, 30, 40, 55, 65, 80, 90, 100 wt%.

## 3. RESULTS AND ANALYSIS OF DATA

### A. Viscosity

The steady dynamic viscosities, $\eta_s^\circ$, measured as a function of temperature at different mass fractions, are displayed in Fig.2a (in Log scale), whereas in Figure 2b they are displayed as a function of the surfactant mass fractions at different temperatures. Data at 1, 2.5 and 5 wt%, taken from Reference 12 are reported too. In Figures 2a and 2b, it possible to distinguish two different concentration regions: a) a dilute and semidilute region (from 0 to about 60 wt%) where $\eta_s^\circ$ exhibits a skewed behaviour, i.e. it first decreases with temperature down to a value $T_{s-}$, then it increases and reaches a maximum at a temperature $T_{s+}$, finally it decreases again (both $T_{s-}$ and $T_{s+}$ decrease with increasing wt%); b) a high concentration region (from 65 to 100 wt%) where $\eta_s^\circ$ decreases monotonically with temperature. Both $T_{s-}$ and $T_{s+}$ are displayed as triangles in Fig. 1. It has to be noted that in the first region $\eta_s^\circ$ increases with wt%, whereas in the second one it decreases (see Fig. 2a-b).

The $\eta_s^\circ$ behaviour in dilute and semidilute regime resembles that already seen in other $C_iE_j$ solutions[5,10,13,14] and microemulsions[15,16,17]. In the latter it was interpreted on the ground of percolation processes and, in analogy, we tried to describe our data according to the scaling relations of the asymptotic behaviour of $\eta_s^\circ$ near a percolation threshold $\phi_c$ [15,16]:

$$\eta_s^\circ = A(\phi - \phi_c)^{\mu'} \quad \text{if} \quad \phi > \phi_c + \delta^+ \tag{1a}$$

and

$$\eta_s^\circ = B(\phi_c - \phi)^{-s'} \quad \text{if} \quad \phi < \phi_c - \delta^- \tag{1b}$$

Although the determination of the scaling exponents of the previous equations is out of the scope of this work, a semiquantitative procedure, used by Peyrelasse et al.[15] in microemulsions, gives reliable $\phi_c$ values. The procedure consists in fitting the experimental Log $\eta_s^\circ$ vs $\phi$ at each temperature by means of a polynomial of an order high enough (fourth or fifth) to well reproduce the experimental data in the whole concentration range. The $\phi_c$ can be then identified with the maxima in the plots [d(Log $\eta_s^\circ$)/d$\phi$] vs $\phi$, corresponding to inflexion points in the Log $\eta_s^\circ$ vs $\phi$ plots. These values are indicated by arrows in Figure 3 and displayed as crosses in the phase diagram in Figure 1. It has to be noted that the line $\phi_c$(T), drown in Figure 1, locate just between the line $\phi(T_{s+})$ and $\phi(T_{s-})$, previously introduced.

The described $\eta_s^\circ$ behaviour agrees with the generally accepted micellar growth with temperature and concentration. Below $T_{s-}$ the viscosity linearly decreases with temperature, indicating that the solution contains aggregates with stable shape and size. Between $T_{s-}$ and $T_{s+}$ $\eta_s^\circ$ is affected by the micellar growth, above $T_{s+}$ a new different regime appears.

At high concentration (from 65 to 100 wt%) the viscosity exhibits a different trend, i.e. it monotonically decreases either with temperature (see Figure 2(a)) and concentration (see Figure 2(b)). The temperature dependence of viscosity can be very well represented with equal accuracy either by the Vogel-Fulcher-Tamman-Hesse (VFTH) equation

$$\eta_s^\circ = A_1 \exp\left[\frac{B}{T - T_1}\right] \tag{2}$$

or by a power-law form

$$\eta_s^\circ = A_2\left[\frac{T}{T_2} - 1\right]^{-m} \quad (3).$$

The previous equations are characteristic of glass-like polymeric systems indicating that the amphiphile-rich samples have the same structural conformation of the pure surfactant, thus it resembles that of polymeric melts, i.e. a glass-like structure.

**B. Sound velocity and longitudinal real modulus**

The behaviours of sound velocity ($c$) as a function of temperature at different concentrations are shown in Figure 4. For all the investigated compositions and temperatures we found no discernible frequency dispersion in the frequency range 5-45 MHz, thus we take the $c$ values in this frequency range as the zero frequency limits. The figure show again two distinct concentration ranges, roughly corresponding to those found for the viscosity: a) a surfactant rich region (from 100% to about 55%) where sound velocity, like the viscosity, regularly decreases either by increasing temperature or the surfactant content; b) a water-rich region (from 0 to about 55%) where it does not follow the trend of $\eta_s^\circ$. In particular at intermediate concentrations 20, 30, 40 wt% the sound velocity does not exhibits any skewing or trend suggestive of percolating process. This behaviour deserves some further considerations, because percolating processes could be masked by the comparable values and by the opposite temperature dependence of the sound velocities of the pure components (see Figure 4). On the contrary, the viscosity of water is always much lower than that of the pure surfactant and exhibit similar temperature dependence (see Fig. 2a).

In such conditions, it is customary to analyse the elastic properties using the excess value of the longitudinal elastic moduli ($M' = \rho c^2$) instead of sound velocity[5,17,18]. The excess properties ($\Delta M' = M' - M'_{ref}$) can be obtained using different reference systems. Usually $M'_{ref}$ is taken as

$$\frac{1}{M'_{ref}} = \frac{1-\phi}{M'_w} + \frac{\phi}{M'_s} \qquad (4)$$

where $M'_w$, and $M'_s$ are the longitudinal moduli of pure water and surfactant respectively. By using equation Eq. 4, known as the Wood relation, the reference medium is an ideal solution, where a volume weighted mixing of the compressibility of the pure components gives the adiabatic compressibility, $\beta_{ref} (= 1/M'_{ref})$. With this choice, $\Delta M'$ represents the excess elastic contribution associated with the interactions between the two components[19]. Unlike the shear viscosity, the calculated $\Delta M'$ values as a function of temperature do not exhibit any evidence of scaling behaviour.

It was recently shown that the volumetric properties of nonionic micellar solutions can be treated as ideal mixture of pure water and hydrated monomer up to 40-50 wt%[6]. Hence the asymmetric solution model, largely used in the literature[20], better represents these mixtures. In this approach the solute–solvent interactions are accounted by attributing all the excess properties only to the surfactant aggregates and the longitudinal modulus $M'$ of solutions is given by:

$$\frac{1}{M'} = \frac{1-\phi}{M'_w} + \frac{\phi}{M'_m} \qquad (5)$$

where $M'_m$ is the “apparent” modulus of the surfactant monomers and $M'_w$ the modulus of pure water. The behaviour of $M'_m$ (calculated by means of Eq. 5) as a function of concentration and temperature is reported in Figure 5 a and b respectively. The figures show that in the water-rich region the surfactant monomers are very rigid (see the peaks of the apparent moduli at around 20 wt%). As the temperature increases, $M'_m$ regularly decreases, i.e. the monomers become softer. For temperatures T>50°C $M'_m$ it takes values even lower than those of the pure surfactant.

In conclusion, we find that, contrary to viscosity, the elastic properties of solutions below 40 wt% do not exhibit any percolation trends. Rather, they seem to be related to the rigidity of the surfactant monomers, which is extremely high at the lowest temperatures and decreases with increasing T. At high concentrations, the elastic modulus shows a regular trend as a function of temperature and concentration similar to that found for $\eta_s^\circ$ and typical of glass-like polymeric melts.

**C. Sound attenuation**

The frequency dependences of the measured sound velocity $c(\omega)$ and amplitude attenuation coefficient $\alpha(\omega)$ are related to the complex longitudinal modulus of the medium $\mathbf{M}(\omega)=\mathbf{K}(\omega)+4/3\ \mathbf{G}(\omega)$ that contains both compressional (**K**) and shear (**G**) components[21,22]. Alternatively, the two quantities can be expressed in terms of the longitudinal complex kinematics viscosity $\mathbf{D}_l(\omega)$[23] given by $\mathbf{D}_l = \rho^{-1}[\boldsymbol{\eta}_v(\omega) + 4/3\boldsymbol{\eta}_s(\omega)]$ that contains complex volume and shear viscosities. The relationships between moduli and viscosities are given by[21]

$$\mathbf{K} = K'(\omega) + \mathbf{i}K''(\omega) = (K_o + K'_r(\omega)) + \mathbf{i}K''_r(\omega) = K_0 + \mathbf{i}\omega\boldsymbol{\eta}_v(\omega) \qquad (6)$$

and

$$\mathbf{G} = G'(\omega) + \mathbf{i}G''(\omega) = \mathbf{i}\omega\boldsymbol{\eta}_s(\omega) \qquad (7)$$

The sound velocity, $c(\omega)$, and the attenuation coefficient, $\alpha(\omega)$, can be related to the dynamic moduli or to the dynamic viscosities by the following equations[21]:

$$c^2(\omega) = \rho^{-1}\left[K_0 + K'_r(\omega) + \frac{4}{3}G'(\omega)\right] \qquad (8)$$

$$\alpha / f^2 = \frac{2\pi^2}{\rho c^3}\left[\frac{K''(\omega)}{\omega} + \frac{4}{3}\frac{G''(\omega)}{\omega}\right] = \frac{2\pi^2}{\rho c^3}\left[\eta_v(\omega) + \frac{4}{3}\eta_s(\omega)\right] = \frac{2\pi^2}{\rho c^3}\eta_l(\omega) \qquad (9)$$

where $f(=\omega/2\pi)$ is the frequency and $\eta_l(\omega)$ the longitudinal viscosity. In the limit of very low (0) and very high (∞) frequencies, Eq. 8 and 9 become

$$c_\circ^2 = \frac{K_\circ}{\rho} \qquad (8')$$

$$(\alpha / f^2)_\circ = \frac{2\pi^2}{\rho c_\circ{}^3}\left[\eta_v^\circ + \frac{4}{3}\eta_s^\circ\right] \tag{9'}$$

and

$$c_\infty{}^2 = c_\circ{}^2 + \frac{1}{\rho}\left(K_r + \frac{4}{3}G_\infty\right) \tag{8''}$$

$$(\alpha / f^2)_\infty \rightarrow 0 \tag{9''}$$

where $\eta_v^\circ(\omega)$ and $\eta_s^\circ(\omega)$ are the static volume and shear viscosity, respectively. In these limits $c$ and $\alpha / f^2$ become independent of frequency.

In ordinary molecular fluids (low $\eta_s^\circ$) the relaxation times of the shear viscosity processes are very short ($\tau_S \approx 10^{-9}$ s) while those of the volume viscosity ($\tau_V$) can be comparable with the characteristic ultrasonic times ($f^{-1} \approx 10^{-6}$ to $10^{-9}$ s), thus Eq. 8 and 9 become

$$c^2(\omega) = c_\circ{}^2 + \frac{1}{\rho}(K_r(\omega) + \frac{4}{3}G_\infty) \tag{10}$$

and

$$\alpha(\omega)/ f^2 = \frac{2\pi^2}{\rho c^3}[\eta_V(\omega) + \frac{4}{3}\eta_S^\circ] = (\alpha(\omega)/ f^2)_V + (\alpha / f^2)_{NS} \tag{11}$$

where

$$(\alpha / f^2)_{NS} = \frac{8\pi^2}{\rho c^3}\eta_s^\circ \tag{12}$$

is the so named Navier-Stokes contribution, that can be evaluated by independent measurements of sound velocity and shear viscosity $\eta_s^\circ(\omega)$. In this case, the frequency dependence of the sound velocity and of the attenuation $\alpha / f^2$ is related to the relaxation of the volume viscosity (e.g. structural, thermal, chemical, or due to other processes[21,24]). Hence in ordinary liquids the experimental $\alpha/f^2$ is never lower than $(\alpha / f^2)_{NS}$.

On the other hand, in viscoelastic media (e.g. complex systems such as microemulsions, micellar phases, glass-forming systems, percolating network, polymer melts and solutions, ..) the relaxation times of both viscosities, volume and shear, are comparable or longer than the characteristic ultrasonic times scale. Thus the observed relaxation can be due to both volume and shear viscosity.

An experimental parameter particularly suited to distinguish ordinary liquids from viscoelastic media, is the scaled attenuation *r*, i.e. the ratio between the experimental $\alpha/f^2$ values and the Navier-Stokes contributions $(\alpha / f^2)_{NS}$ given by 12. Values of *r* bigger than one correspond to ordinary molecular liquids, lower than one to viscoelastic media.

Since both $\eta_s^\circ(\omega)$ and $M'_m$ indicate a glass like structure in the surfactant rich region, we analyse our attenuations by means of the reduced quantity *r*, a clear indicator of viscoelesticity.

In Figures 6 a-b we plotted the scaled attenuation $r(f, wt\%, t) = (\alpha / f^2)_{\exp} / (\alpha / f^2)_{NS}$ as a function of frequency (*f*) for all the investigated concentrations and temperatures[25]. In most of the samples the ratio *r(f)* becomes smaller than 1, a clear evidence of viscoelasticity. An overall view of Figures 6 shows that the values and the trends of the scaled attenuations behave differently in the surfactant–rich and in the water-rich solutions, therefore we analyse separately the two regions (Figure 6a and 6b).

<u>a) Concentrated solutions</u> (~ 65 to 100wt%)

In these viscous solutions the scaled absorption *r* is slightly higher than 1 at low frequencies and smaller than 1 at high frequencies. This represents a direct evidence of a relaxation of both shear and volume viscosities, with relaxation frequencies in the investigated range. It is worth to note that the *r(f)* curves in Figure 6a shift toward higher frequencies as temperature increases. The fact that the steady shear viscosity of these solutions decreases with increasing temperature (see Figure 2a) suggests a link between characteristic relaxation times $\tau$ of solutions and $\eta_s^\circ$. Such link is indeed well established in the ultrasonic studies of viscous molecular liquids where the steady volume ($\eta_v^\circ$) and shear viscosity ($\eta_s^\circ$) are related to the relaxation times by $\eta_v^\circ = K_v \tau_v$ and $\eta_s^\circ = G_\infty \tau_s$. Following these suggestions, we then describe the dynamics of relaxation in according to the general form of Eq. 9, assuming for both volume and shear contributions the well-known Cole-Cole relaxation formula[24,26]

$$\eta_v(f) = \frac{\eta_v^\circ}{1 + (f / f_v)^{2(1-\beta)}}; \eta_s(f) = \frac{\eta_s^\circ}{1 + (f / f_s)^{2(1-\beta)}} \qquad (13)$$

Eq.13 represent a continuous distribution of processes around two average relaxation frequencies, where $\beta$ is a parameter that defines the distributions of the relaxation processes, with characteristic times $\tau_v = (2\pi f_v)^{-1}$ and $\tau_s = (2\pi f_s)^{-1}$. Since in viscous molecular liquids it is often found $\tau_v \approx \tau_s \approx \tau$, we tentatively assumed this equality so that Eq. 9 reduces to

$$\frac{\alpha}{f^2}(f, wt\%, T) = \frac{A(wt\%, t)}{1 + (f / f_r)^{2(1-\beta)}} + B = \frac{A(wt\%, t)}{1 + (2\pi f \eta_s^\circ / G_\infty)^{2(1-\beta)}} + B \qquad (14)$$

where the amplitude $A = (\alpha / f^2)_\circ = (2\pi^2 / \rho c_\circ^3)[\eta_v^\circ + 4/3\eta_s^\circ] = (\alpha / f^2)_{\circ,V} + (\alpha / f^2)_{NS}$ is the zero frequency attenuation, $f_r = (2\pi\tau)^{-1}$ the common relaxation frequency and B is the limit of absorption at high frequency. The scaled absorption $r(f) = (\alpha / f^2)(\alpha / f^2)_{NS}$, according to Eq.14, is given by

$$r(f) = \frac{\alpha / f^2(f, wt\%, t)}{(\alpha / f^2)_{NS}} = \frac{A^*(wt\%, t)}{1 + (2\pi f \eta_s^\circ / G_\infty)^{2(1-\beta)}} + B^* \qquad (15)$$

where $A^* = A/(\alpha/f^2)_{NS} = 1+[\eta_v^\circ/(4/3)\eta_s^\circ]$ and $B^* = B/(\alpha/f^2)_{NS}$ are reduced quantities. Assuming that the parameters $A^*$, $G_\infty$, $\beta$ and $B^*$ of Eq. 15 do not depend strongly on concentration and temperature, the data points at any temperature and concentration should fall on a single curve if plotted as a function of the scaled frequency $2\pi f\eta_s^\circ/G_\infty$. This can be seen in Figure 7, where *r(f)* points are displayed as a function of the scaling parameter $2\pi f\eta_s^\circ$ that is proportional to the scaled frequency $2\pi f\eta_s^\circ/G_\infty$. The line is the best fit of Eq. 15 to the data and the relative parameters are listed in the Table I. The reliability of the parameters in the given frequency range is supported by the correlation parameter *R*. Since the data points approximately scale on a single curve, the observed relaxations correspond to the high frequency part of a viscoelastic process. Better fits were obtained by considering only data at the same concentration but at different temperatures. The results of these fits are also reported in Table 1. It can be seen that the fit parameters fluctuate around average values nearly equivalent to those obtained using all the scaled data, thus the observed fluctuation must have a statistical origin. This consideration supports the simplifying assumptions made: $\tau_V \approx \tau_S \approx \tau$ and $A^*$, $G_\infty$, $\beta$, $B^*$ independent of concentration.

From the parameters reported in the Table I we can evaluate the ratio of the static volume to shear viscosity $\eta_v^\circ/\eta_s^\circ = 4/3(A^*-1)$ and the relaxing compressibility modulus $K_v = G_\infty\eta_s^\circ/\eta_v^\circ$ (see Table I). The ratio $\eta_s^\circ/\eta_v^\circ$ takes values (from 1 to 2), typical of associated liquids[24], while the values of elastic moduli $G_\infty$ and $K_V$ ($\approx 10^8$ dyne/cm$^2$) are two order of magnitude smaller than those reported for molecular viscous liquids. However, they are of the same order of magnitude as those found in dense microemulsions[26] and other dense micellar solutions. The small values of $K_V$ and $G_\infty$ can explain the lack of dispersion effects we found on sound velocity in the range 5-45 MHz. Indeed, according to the viscoelastic theory (see Eq. 8'),

$$c_\infty{}^2 - c_0{}^2 = \rho^{-1}\left(K_r + \frac{4}{3}G_\infty\right) \tag{16}$$

thus $\Delta c/c_\circ \approx (2\rho c_\circ{}^2)^{-1}(K_v + 4/3G_\infty)$ is of order $10^{-2}$. These values are quite smaller than those typical of viscoelastic molecular liquids where the amplitude of sound velocity dispersion is very large [21].

In conclusion, we find that ultrasonic absorption dynamics in the concentrated solutions is dominated by viscoelastic effects that are strictly linked to the behaviour of static shear viscosity. This finding is typical of glass–like systems.

b) semidiluted and diluted concentrations (40 to 9%)

The behaviour of the scaled absorption *r(f)* at lower concentrations (Figures 6b) is more complex and strongly dependent on the concentration and temperature. Let us start considering the 40 wt% solutions that are located near the hexagonal phase (see Figure 1). From Figures 6b1 we can see that at the lowest temperature (20 and 30 °C) the system exhibits a viscoelastic behavior similar to that present in the high concentrated region. Accordingly, we find that the ratio *r(f)* can be well represented by eq. 15 with fitting parameters (reported in Table 1) that are comparable with those at high concentrations. However, on increasing the temperature from 30 to 40°, the high frequency tail of *r(f)* suddenly falls, indicating a slowing down of the viscoelastic processes. This temperature interval includes the extrapolated value of $T_{s+}$ (see Fig. 1), i.e. the temperature where viscosity has a maximum. Above 40 °C, *r(f)* raises again and takes values much higher than 1 at the lowest frequencies. These findings indicate that above 40 °C ($\approx T_{s+}$) a new, low–frequency, volume relaxation process superimposes to the high-frequency viscoelastic one.

At lower concentrations (30, 20 and 9 wt%), we can see (Figure 6b2, 3 and 4) that the high frequency viscoelastic effects fades away and the dynamic is dominated by the new low-frequency processes already observed at 40 wt%. However, it must be noted that the inversion in the temperature dependence of *r(f)*, observed at 40 wt%, is still present at lower concentrations. At 30 and 20 wt% it occurs between 40 and 50 °C and between 50 and 60 °C respectively. At 9 wt% the trend suggests that it occurs above 60 °C. Again these temperatures nearly correspond to $T_{s+}$, the $\eta_s^{\circ}$ maxima in Figure 2a.

The above peculiar trends of *r(f)* in the dilute and semidilute regime are suggestive of a strict relationship between the high-frequency tail of ultrasonic dynamics and the steady shear viscosity, as found at the high concentrations. However, the presence of the low-frequency high-amplitude volume relaxation process, and the gradual reduction of viscoelastic effects on decreasing concentration, are characteristic features of water-rich solutions.

In conclusion, we find that the ultrasonic dynamics in dilute and semidilute solutions is, in general, characterized by a superposition of an high-frequency viscoelastic relaxation (still related to the steady shear viscosity) and a low-frequency volume viscosity relaxation, that becomes dominant as the temperature increases and concentration decreases. Large volume viscosity relaxation processes have been frequently detected in very diluted micellar solutions[27] and attributed to exchange processes of surfactant and/or water molecules.

## 4. Discussions and Conclusions

In agreement with previous studies, our analysis of static and dynamic properties on the $C_{12}E_8$ nonionic surfactant solutions supports the existence of two distinct concentration ranges, characterized by different structural conditions. In order to gain more physical insights from our experiments, we follow the evolution of the system as the concentration of surfactant decreases.

The structure formed by the pure $C_{12}E_8$ in its liquid state is characterized by two distinct regions, one incorporating the hydrophilic heads and the other the hydrophobic tails[11]. In the resulting bicontinuous structures the shear viscosity and the ultrasonic dynamics follows eq. 2 and eq 15 respectively. This is characteristic of viscoelastic behaviour generally present in glass-like systems.

On moving from the pure surfactant liquid state to concentrated aqueous solutions (from 100%w to about 55-65%w), the viscosity and the attenuation of the system decrease on increasing temperature still following eq. 2 and 15 respectively. Thus the structure of the pure component is retained in these rich-surfactant solutions.

The role of the water molecules that hydrates the surfactants can be understood by considering that, at fixed temperature, the viscosity and the rigidity of the system increase with the concentration of the solvent (see Fig.2b, 4 and 5b). This behaviour indicates that the water added to surfactant promotes the aggregation between surfactant monomers, enhancing the viscosity and the rigidity of the system. A probable origin of such effect might be the bridging of two or more surfactant monomer by means of H-bonds between water molecules and hydrophilic EO groups belonging to two distinct monomers, as results from recent molecular dynamics simulations[28]. This bridging mechanism is reinforced by the presence of the hydrophobic surfactant tail. The constraint imposed by the confinement of the tail in the core of the micelle (i.e. the anchorage to the hydrophilic-hydrophobic interface)[11] forces the EO groups to be closer to each other and to assume a more extended conformation[29]. Both conditions better expose the EO groups to the external water.

The role of hydrophobic tails in promoting H-bound of water with hydrophilic EO polymeric heads is further supported by the comparison with the experimental trend of viscosity and rigidity of concentrated aqueous solutions of PEG400[30] and PEG600[31]. These polymers are composed by a number (j=8 to 13) of hydrophilic (EO) units, comparable with that in $C_{12}E_8$. When the temperature is varied, the viscosity of these polymeric solutions follows the VFTH equation (Eq.2), like $C_{12}E_8$. On the other hand the viscosity and the rigidity of the polymers decrease on reducing concentration, in contrast with $C_{12}E_8$. The geometrical constraint imposed by the hydrophobic tails could favour the H-bonding between EO segments of different surfactant molecules, whereas PEG polymer would tend to have mainly intra-chain H bonds. This is supported also from molecular dynamic simulations[32,33].

The ultrasonic dynamic observed at high surfactant concentration resembles that shown by many short–chained linear polymer melts (6 to 10 units), such as polysyloxanes[24], polystyrene[34] and polyethylene glycols PEO[35,36], as well as that shown by concentrated aqueous solutions of these molecules. A detailed discussion of the dynamical processes originating these viscoelastic spectra is outside the aim of this paper since a large literature already exists on the subject. In general theories and experiments suggest that the underlying mechanisms are usually associated either to high frequency (> ~3 MHz) conformational and/or local motions of the polymer segments (Rouse-Zimm modes) or to entanglement (low frequency modes) (see for example Ref. [37]). Since the properties of the hydrophobic core are quite insensible to the amount of water in solution and the relaxation frequencies in our experiments depend on water content, we ascribe this dynamic mainly to the hydrophilic terminations. Therefore, the observed inversions of the scaled attenuation *r(f)* with temperature can be related to the well known dehydration process occurring at the PEO-water interface.

The crossover concentration (about 60-65%w), separating the glass-like from the dilute and semidilute regions, can be also explained on the ground of H-bounds network: as long as binding sites are available, the water molecules bridge the EO terminations becoming part of the monomers. As soon as the amount of water molecules reaches the upper limit, corresponding to the saturation of their H-bonds with the available EO groups, a new regime occurs where hydrated surfactant aggregates coexist with free water. Assuming that at most two water molecules can bind each EO unit, besides three more H-bounds can be made with the O-H termination, and taking into account that $C_{12}E_8$ has eight EO units, the saturation should occur at a mole fraction surfactant concentration (x) fulfilling the condition (1-x)/x=19, corresponding to wt%=0.62. This concentration should represent the lower limit for the existence of glass-like solutions and it is close to the concentration where we identify the crossover between water-rich and surfactant-rich solutions in all our experiments. Similar conclusions were found for PEG600[31].

The high frequency tail of ultrasonic dynamics observed in the water rich solutions occurs in the same frequency range of those at high surfactant concentrations, besides it presents the same inversions of the scaled attenuation *r(f)* with temperature. Thus we still associate this dynamic with the conformational and/or local motions of the hydrophilic portion of surfactant monomers. As a matter of fact, there are evidences that the EO-water mixture at the micellar interface resembles that of a high concentrated PEO-water solution[6]. It must be mentioned that the inversions of the *r(f)* temperature dependence occur at temperatures close to those relative to the maxima in the viscosity, $T_{s+}$ (see fig 2a). These temperatures define the crossover between a solution of dispersed cylindrical aggregates and one composed by entangled micelles[5]. Above $T_{s+}$ these solutions acquire a glass-like behavior, as generally expected for entangled polymer mixtures. But in the case of micelle the entanglement implies that the hydrophilic terminations of different aggregates are nearly in contact, a condition resembling the one of surfactant rich solutions.

In the water rich region, we observed the incoming of a low-frequency tail in the ultrasound attenuation, whose amplitude increases as the concentration of the surfactant decreases. This dynamical contribution is likely due to the water and/or monomers exchange between the aggregates interface and the free water region[6].

The behaviors of the viscosity and the rigidity of the solution at high water content reflect the shape and size of the micellar aggregate. In the region below $T_{s-}$ just spherical aggregates are present, thus the viscosity decreases only for thermal effects. The same can be stated for the rigidity. Between $T_{s-}$ and $T_{s+}$, the viscosity increases because the micelle growth. The rigidity increases also at low temperature, indicating the presence of stiff aggregates but, it decreases to values even smaller than those observed in the pure surfactant when the temperature increases. This behaviour must be related to the significant decreases with temperature of the free energy gain on forming micellar aggregates[38].

In conclusion from our experiments we can state that two concentration regions characterize the non-ionic micellar solutions, the boundary between the two is a line in the temperature-concentration phase diagram identified by $T_{s+}$. In the water rich region the average properties of the micellar aggregates and the exchange of monomers and/or hydration water molecules determine the experimental results. In the surfactant rich region viscoelasticity and glass-like properties govern the behavior of the solutions. Clearly the structural deductions reported above are limited by the techniques used, and spectroscopic information (nuclear magnetic resonance and/or vibrational) are needed to confirm them. This will be object of future work.

**ACKNOWLEDGEMENTS**
The financial support of the Istituto Nazionale di Fisica della Materia (INFM), Italy, is gratefully acknowledged.

**FIGURES AND TABLES CAPTIONS**

**Figure 1**. The phase diagram of the water-$C_{12}E_8$ system (from Ref. [1]). Measurements were performed in the isotropic phase $L_1$. Dashed line is the percolation line obtained from the shear viscosity data as discussed in the text. Lower and upper triangles are the points where $\eta_s^\circ(T)$ curves in Fig.2a exhibit minima ($T_{s-}$)and maxima ($T_{s+}$)

**Figure 2**. The temperature (a) and concentration (b) dependence of the measured shear viscosity $\eta_s^\circ$ of solutions. Data at 1, 2.5 and 5 wt% are from Ref. [9]. Lines are smooth curves through the experimental data.

**Figure 3** The derivative of Log $\eta_s^\circ$ with respect to volume fraction $\phi$ (≈wt%) as a function of wt% at various temperatures. Log $\eta_s^\circ(\phi)$ are 4-5 degree polynomials fitting the experimental $\eta_s^\circ(\phi)$ data in Figure 2(b). Arrows indicate the locations of the maximum of each curve and correspond to the inflection points ($\phi_c$) of Log $\eta_s^\circ(\phi)$ curves in Figure 2(b). $\phi_c$ values are displayed in Figure 1.

**Figure 4**. The temperature dependence of the experimental sound velocity at 15 MHz. Lines are guides to the eyes.

**Figure 5** The concentration (a) and temperature (b) dependence of the "apparent" real longitudinal modulus $M'_m$ of the surfactant. $M'_m$ is calculated from the moduli of solutions ($M'$) and of the water ($M'_w$) by using Eq.5. Lines are guides to the eyes.

**Figure 6**. The scaled attenuation $r\left[= (\alpha / f^2)_{\exp} / (\alpha / f^2)_{NS}\right]$ as a function of frequency *f*. Figure a: 100, 90, 80 and 65 wt%. Figure b: 40, 30, 20, 9 wt%.

**Figure 7.** The scaled attenuation *r* as a function of $f\eta_s^\circ$ at high concentrations (100, 90, 80, 65 wt%) and all the temperatures. The line is the best-fitting Cole-Cole eq.15 whose parameters are reported in Table 1.

**Table 1**. Best fitting parameters of the Cole-Cole Eq. (15) describing the scaled ultrasonic absorption in the rich-surfactant concentrations region.(see Fig. 7) .

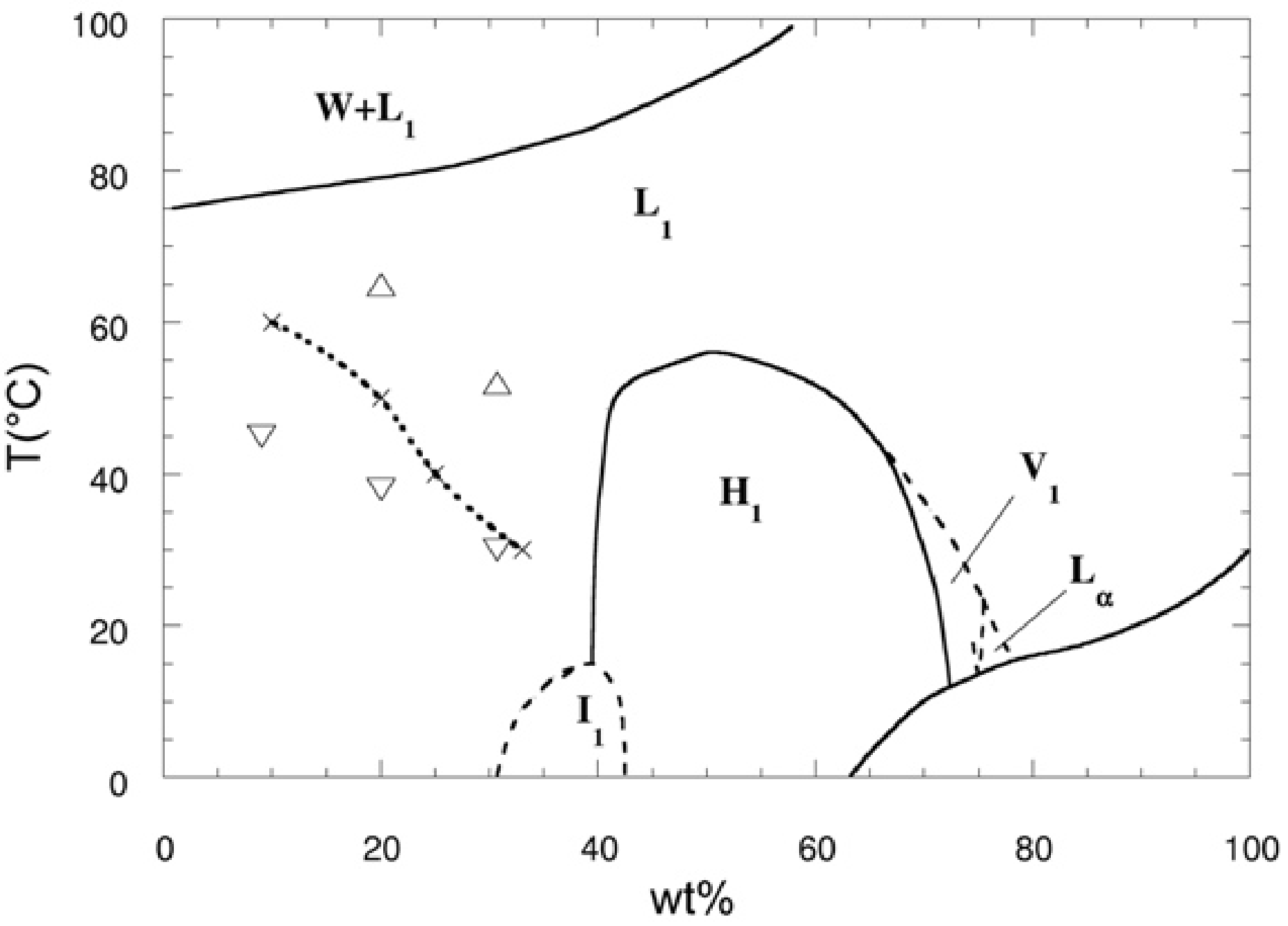

Figure 1

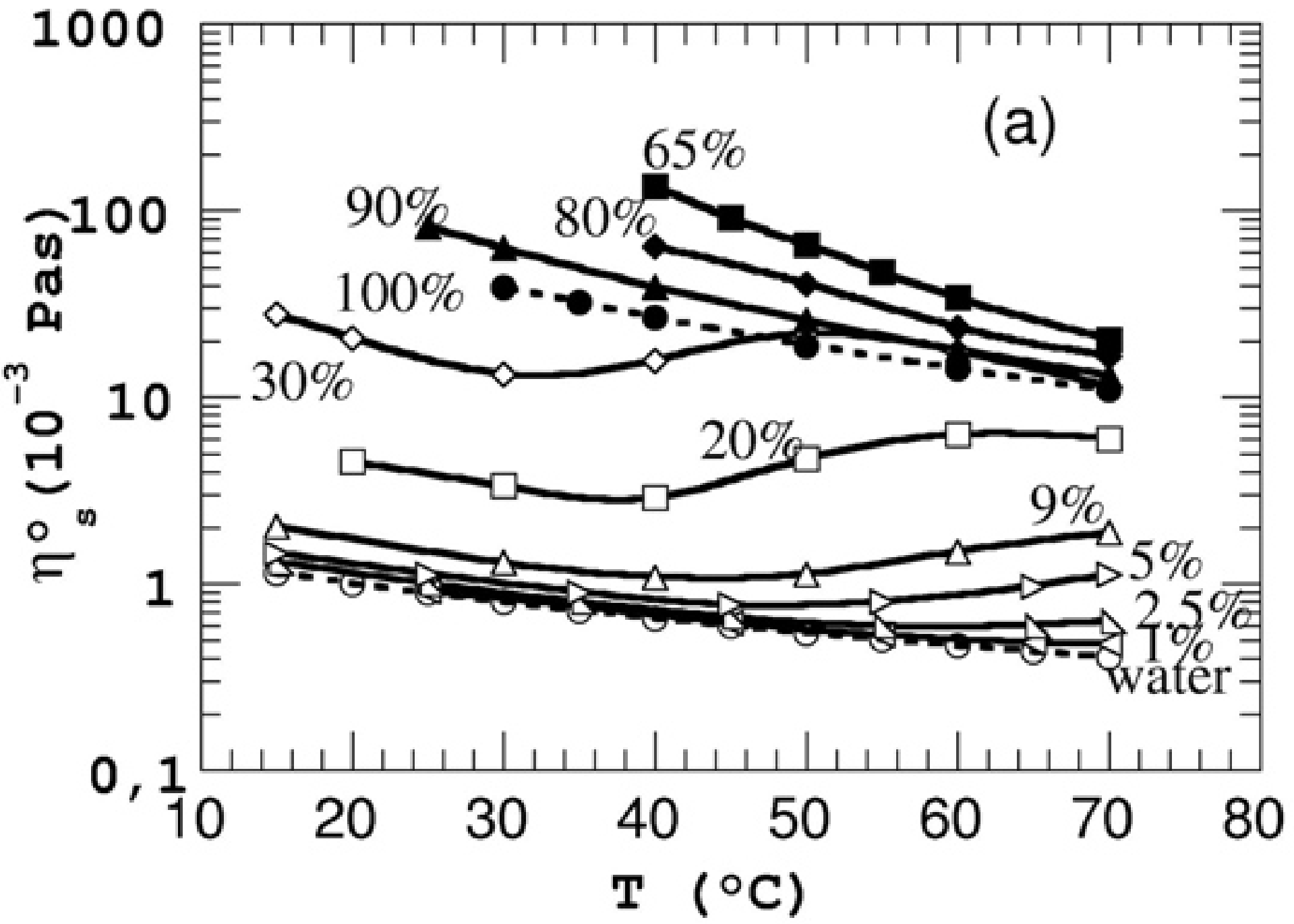

Figure 2a

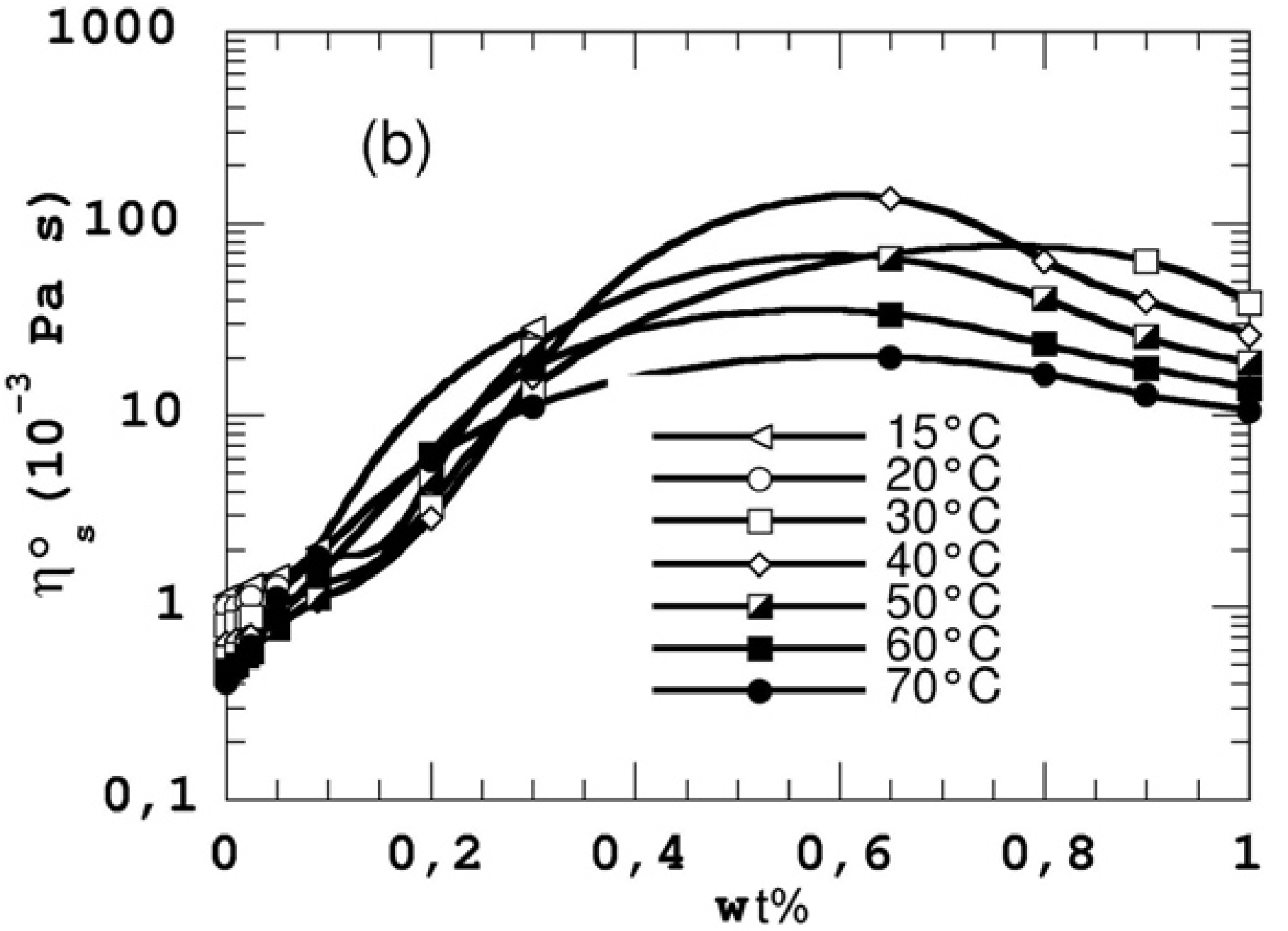

Figure 2b

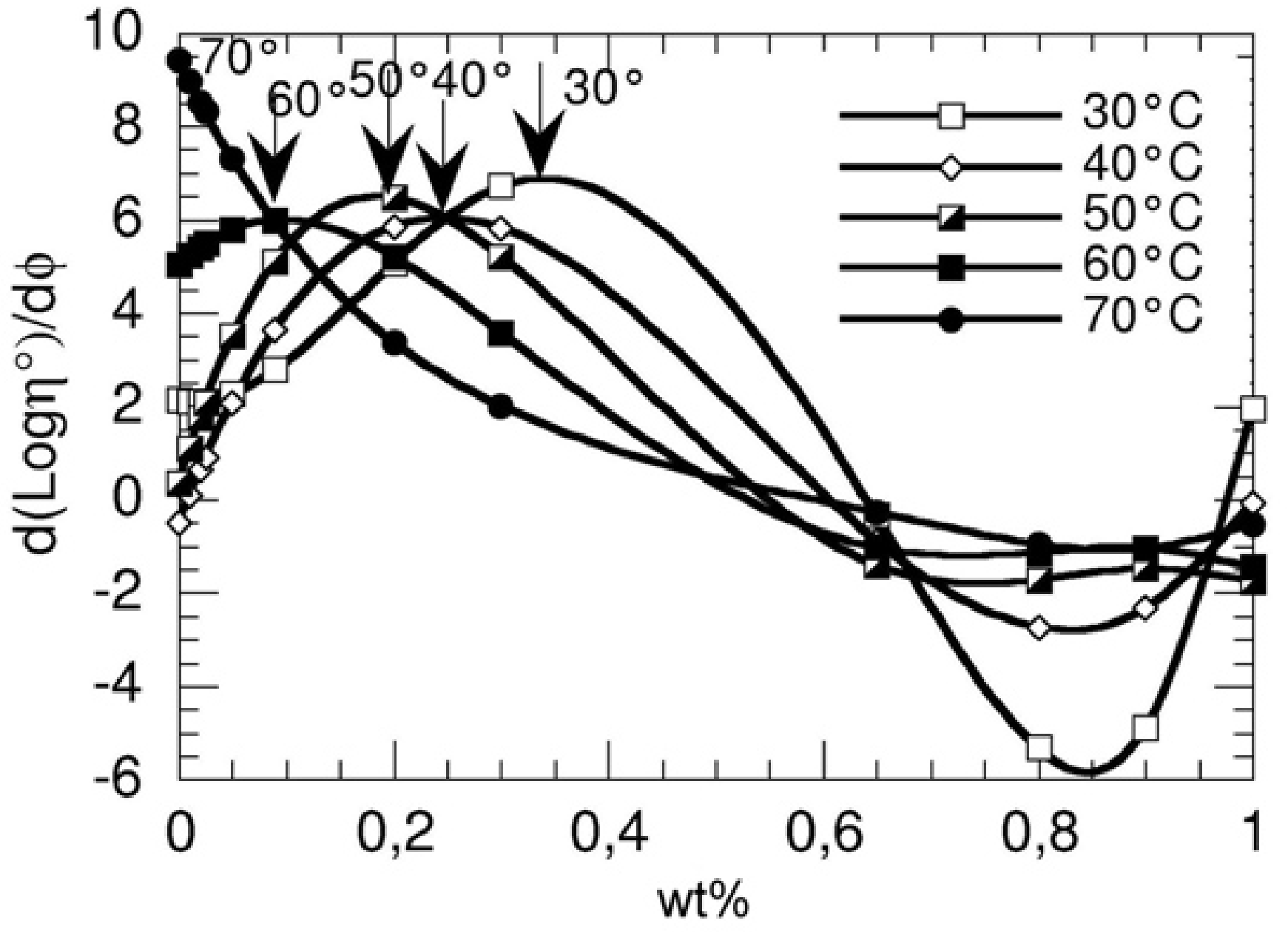

Figure 3

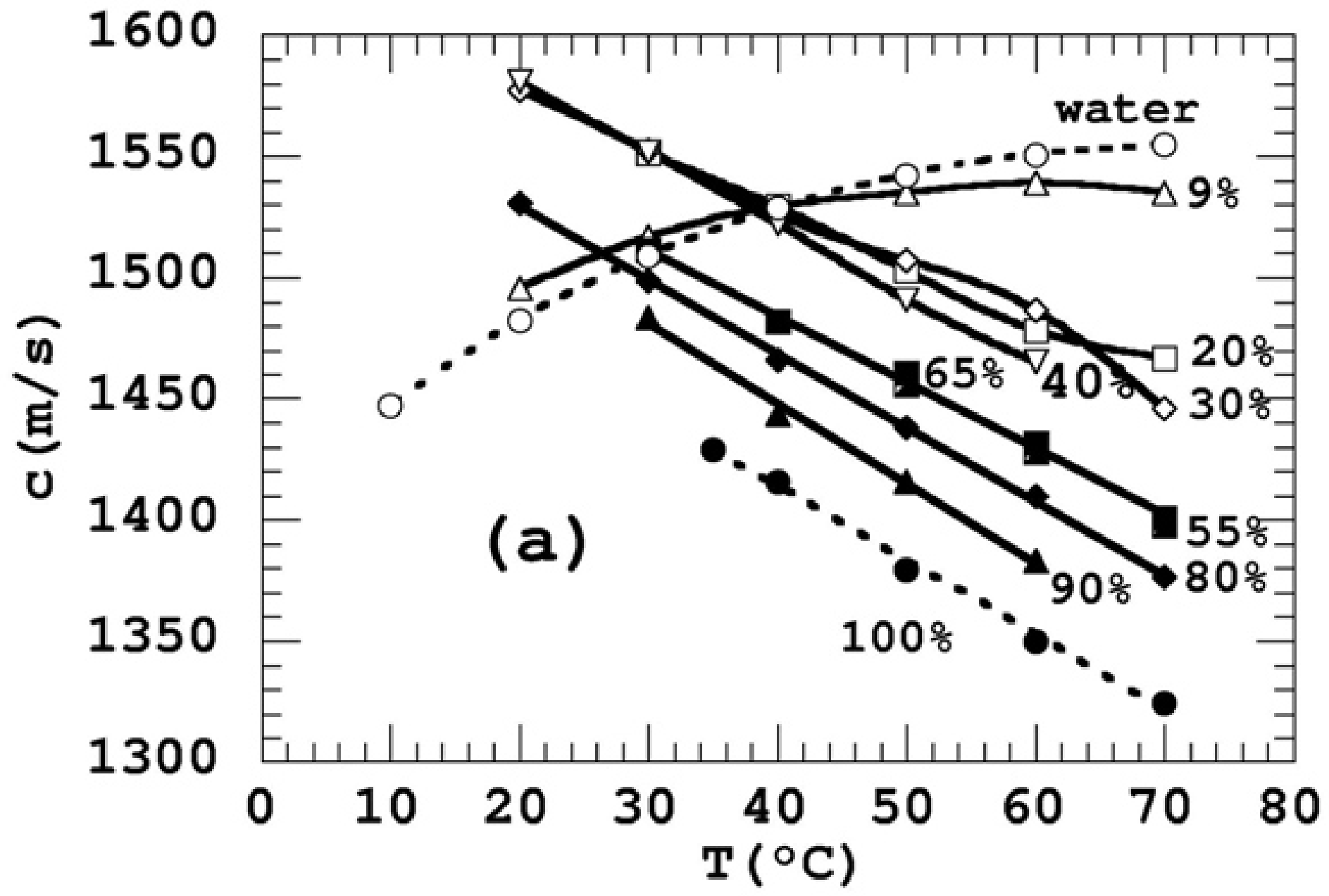

Figure 4

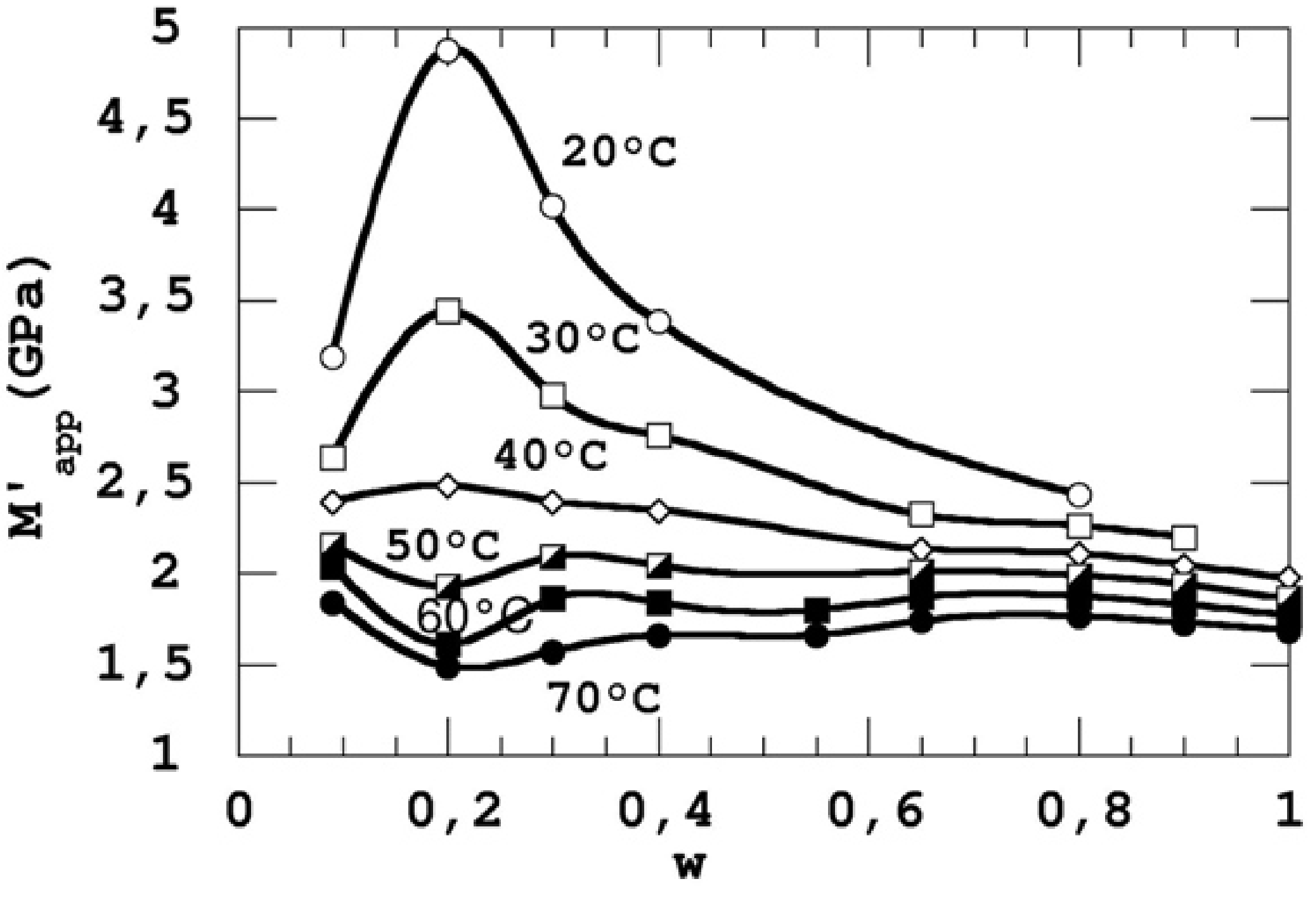

Figure 5a

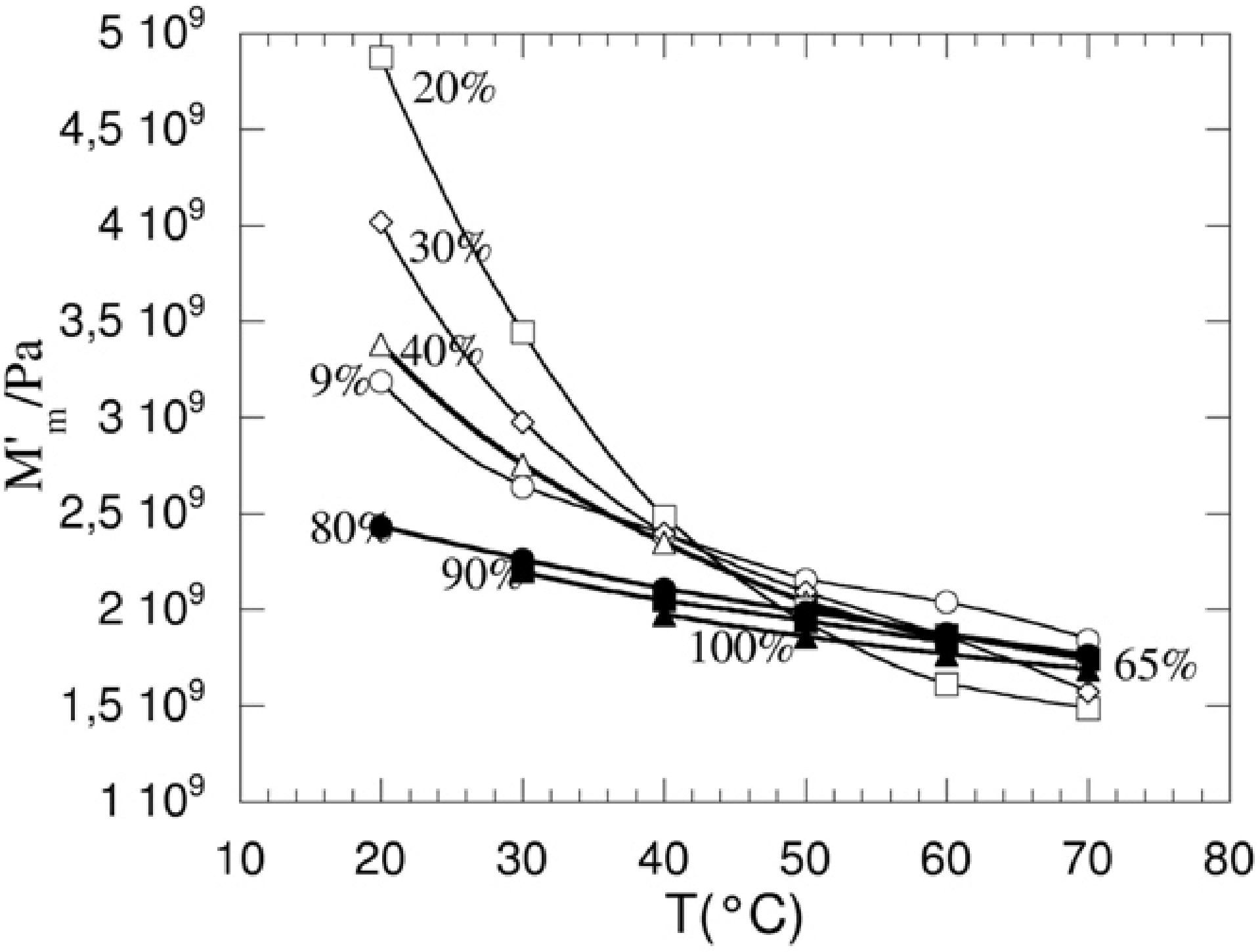

Figure 5b

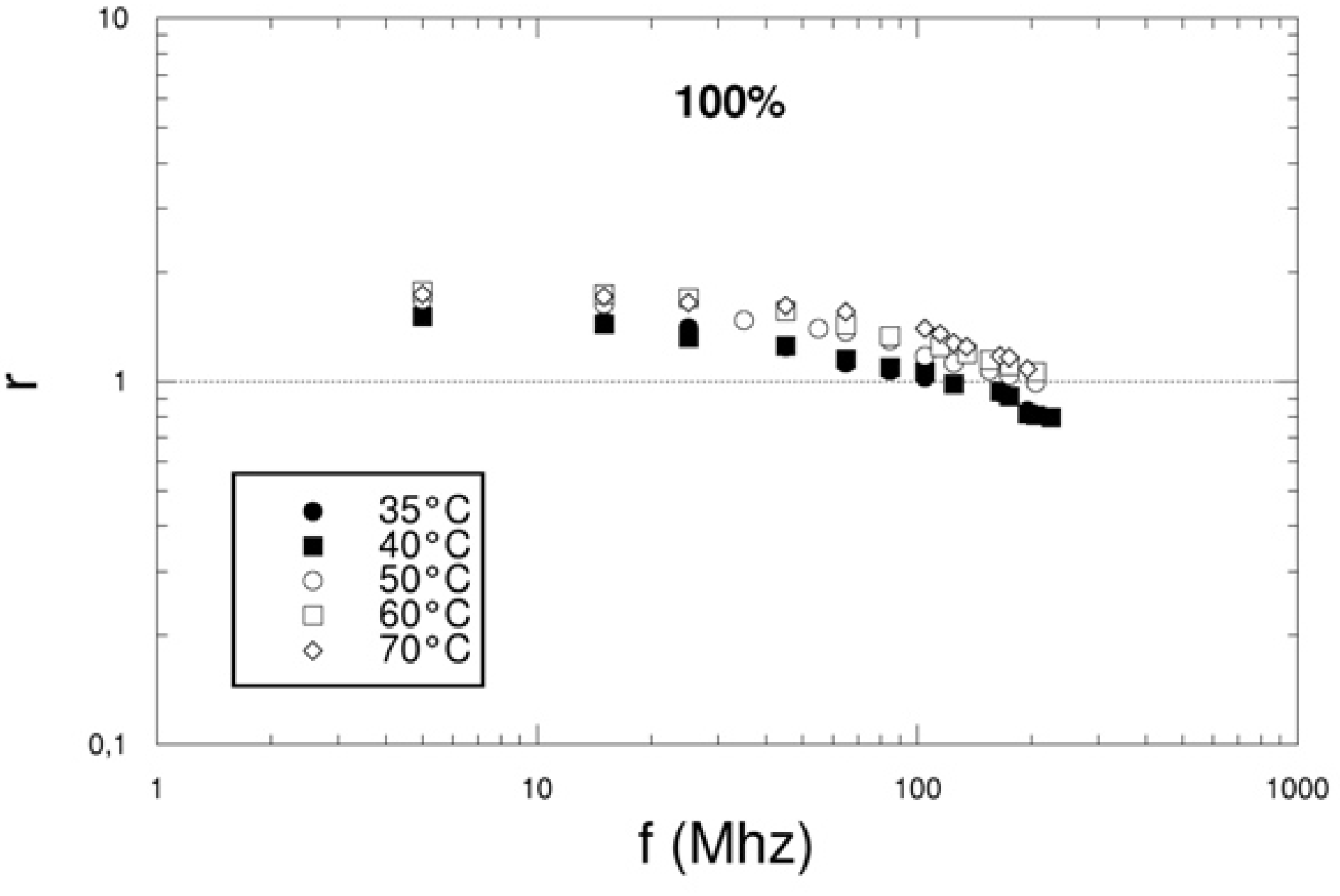

Figure6a1

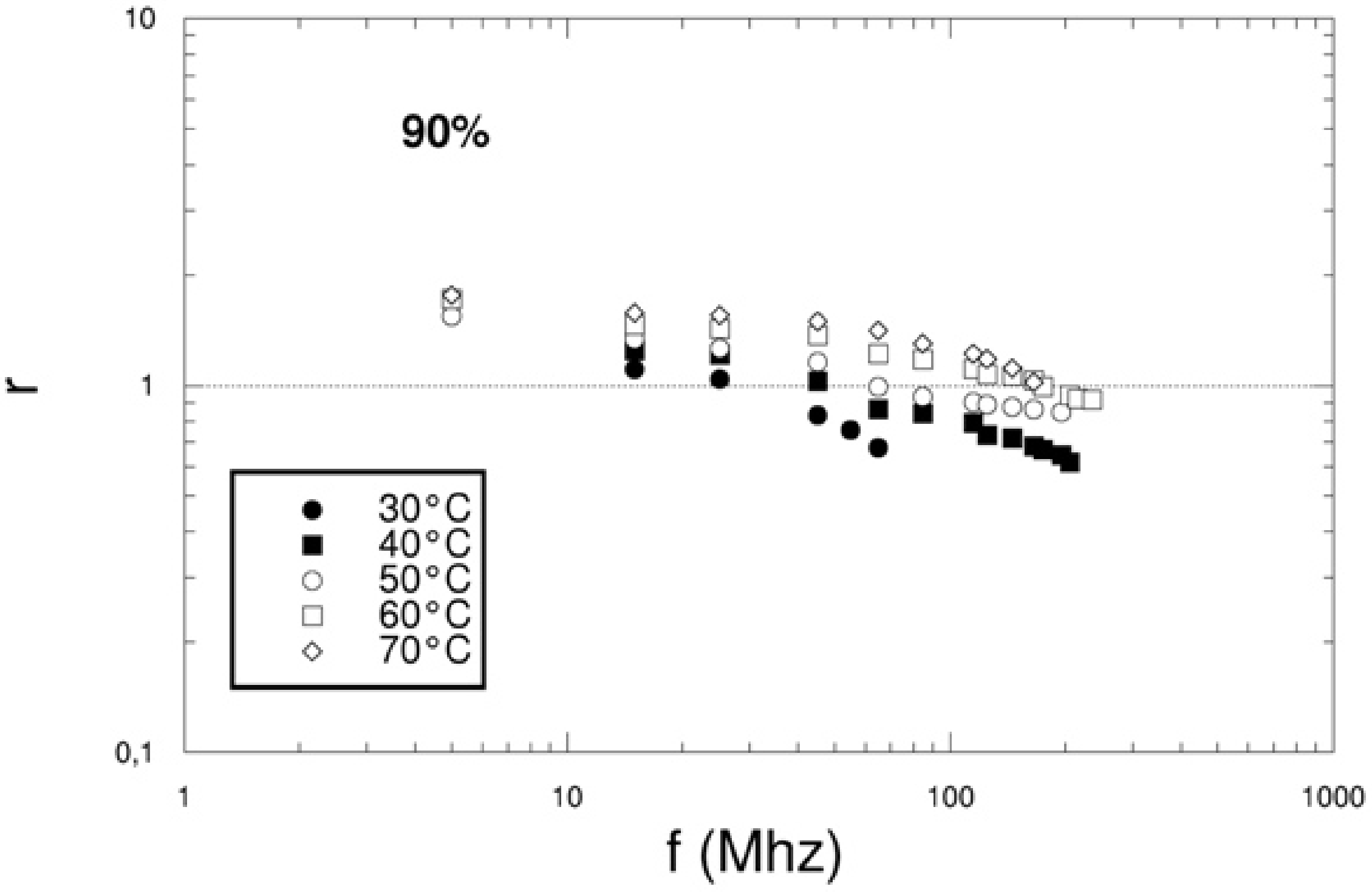

Figure 6a2

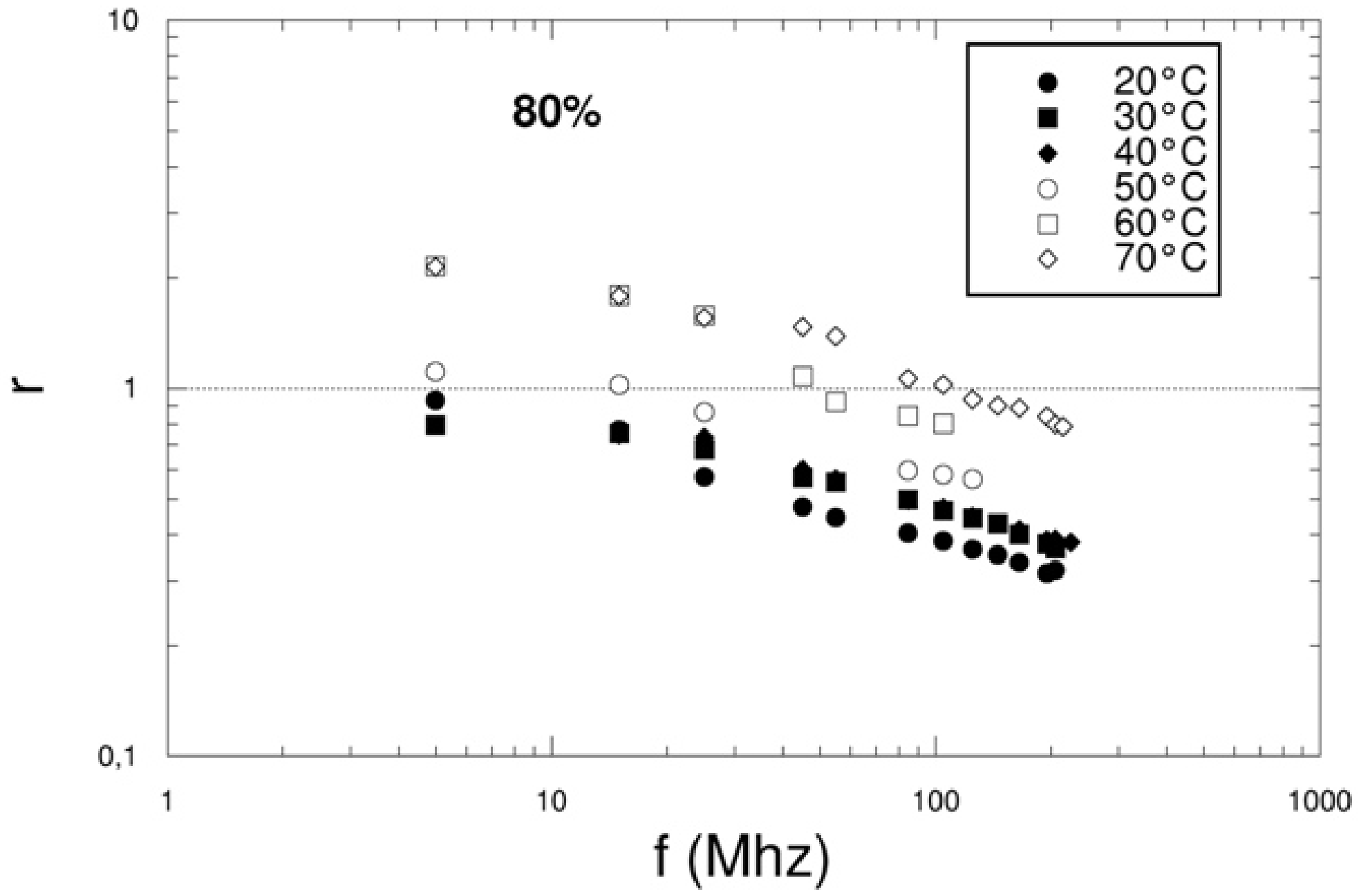

Figure 6a3

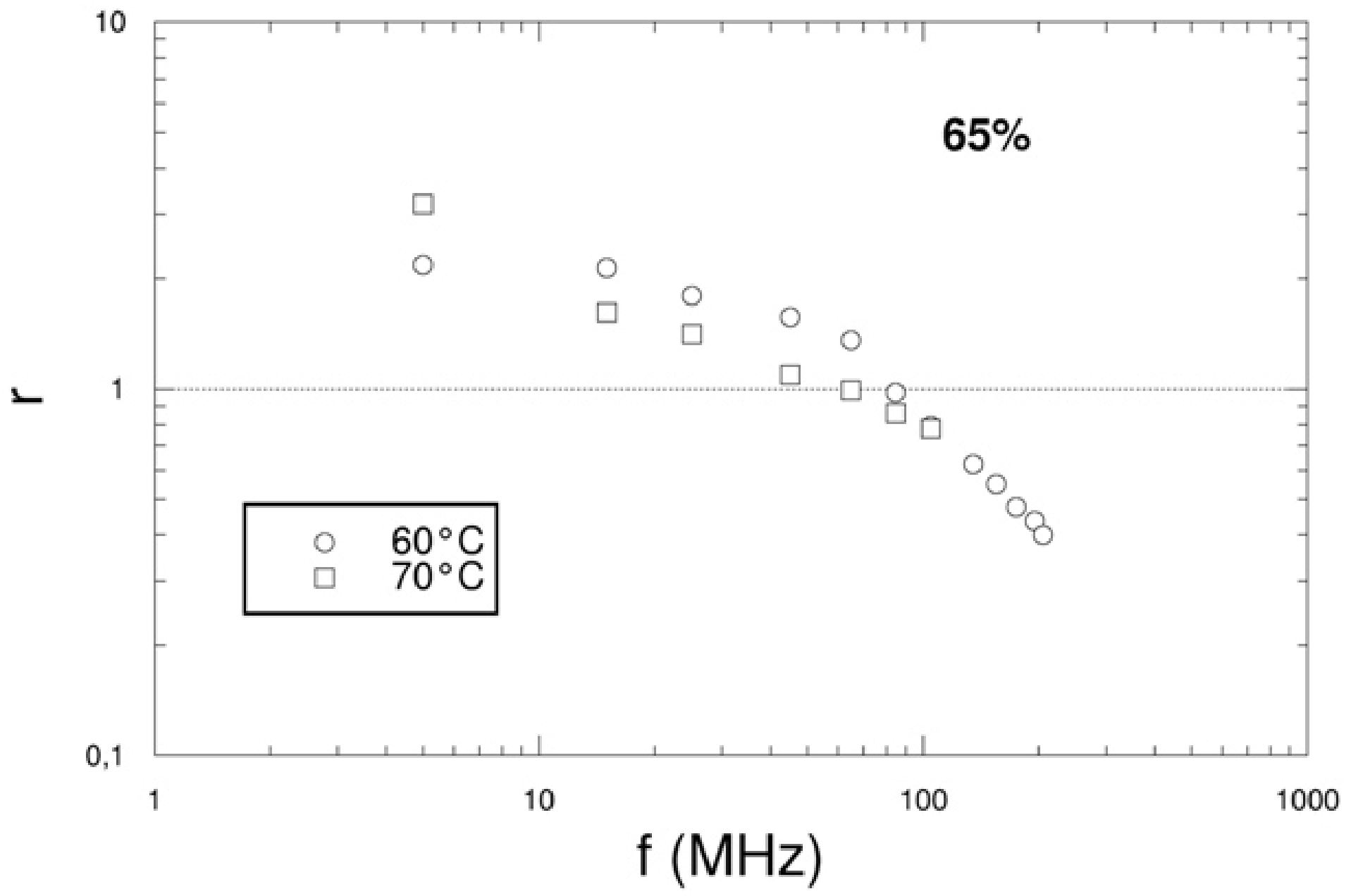

Figure 6a4

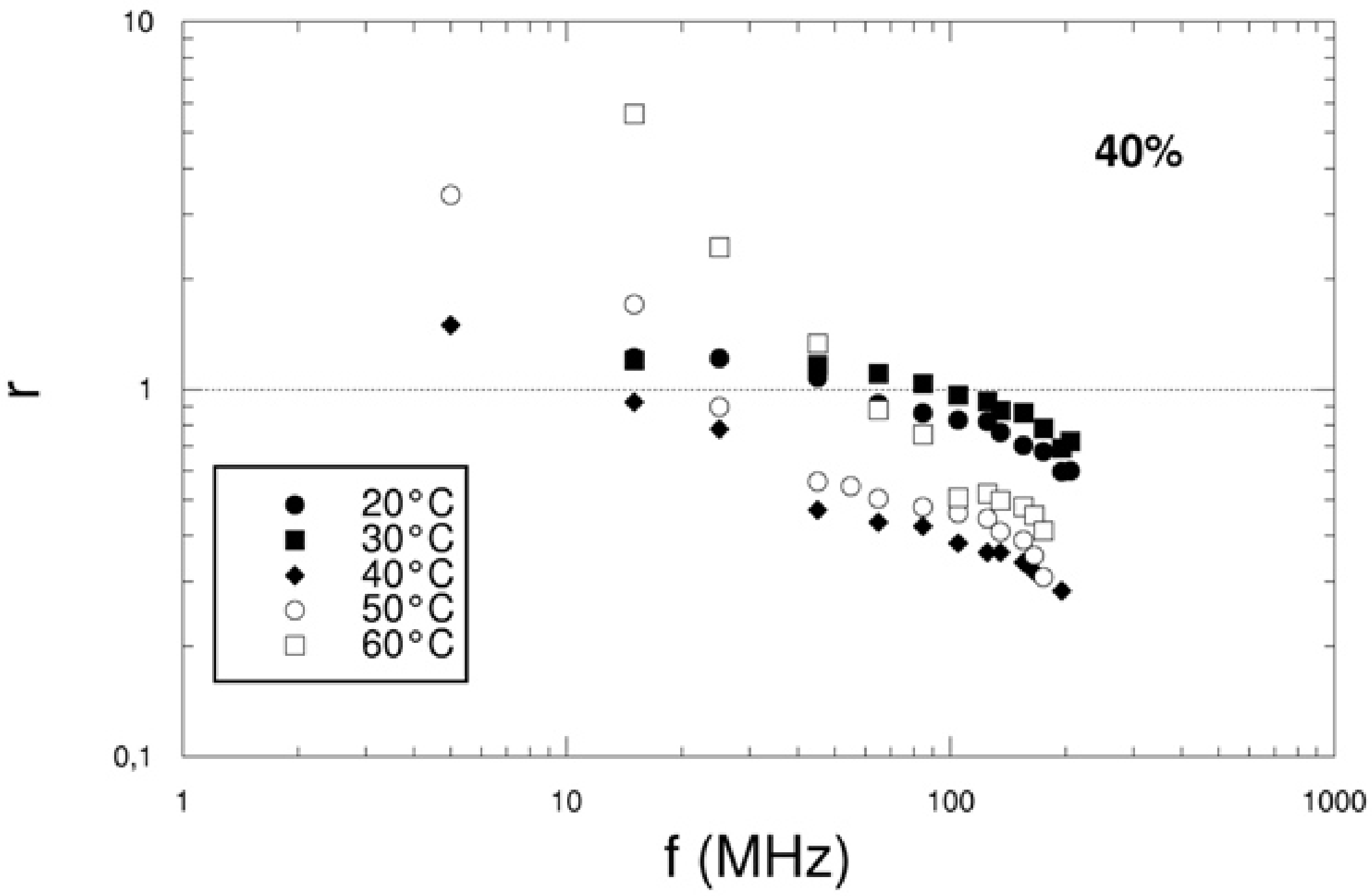

Figure 6b1

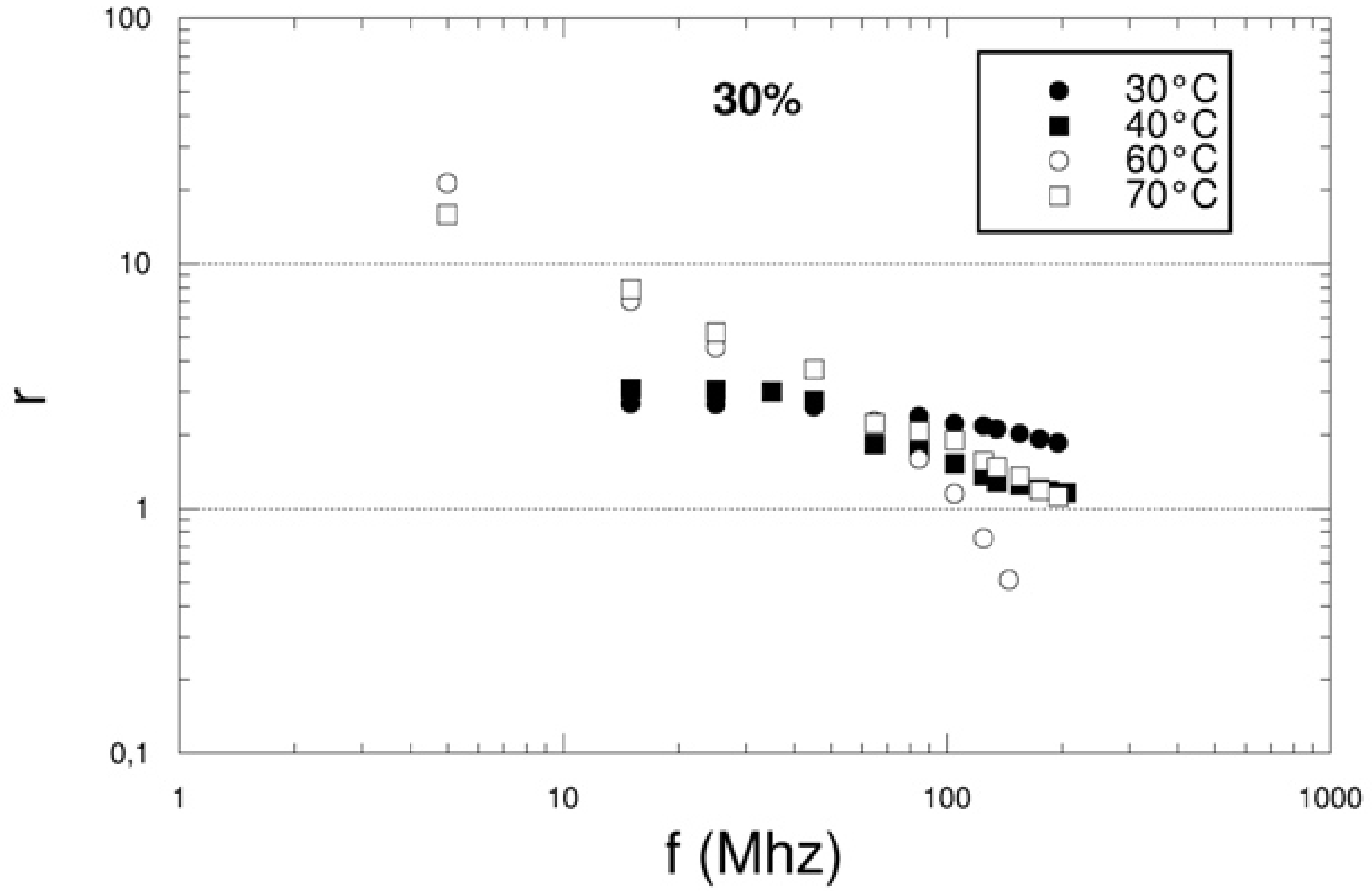

Figure 6b2

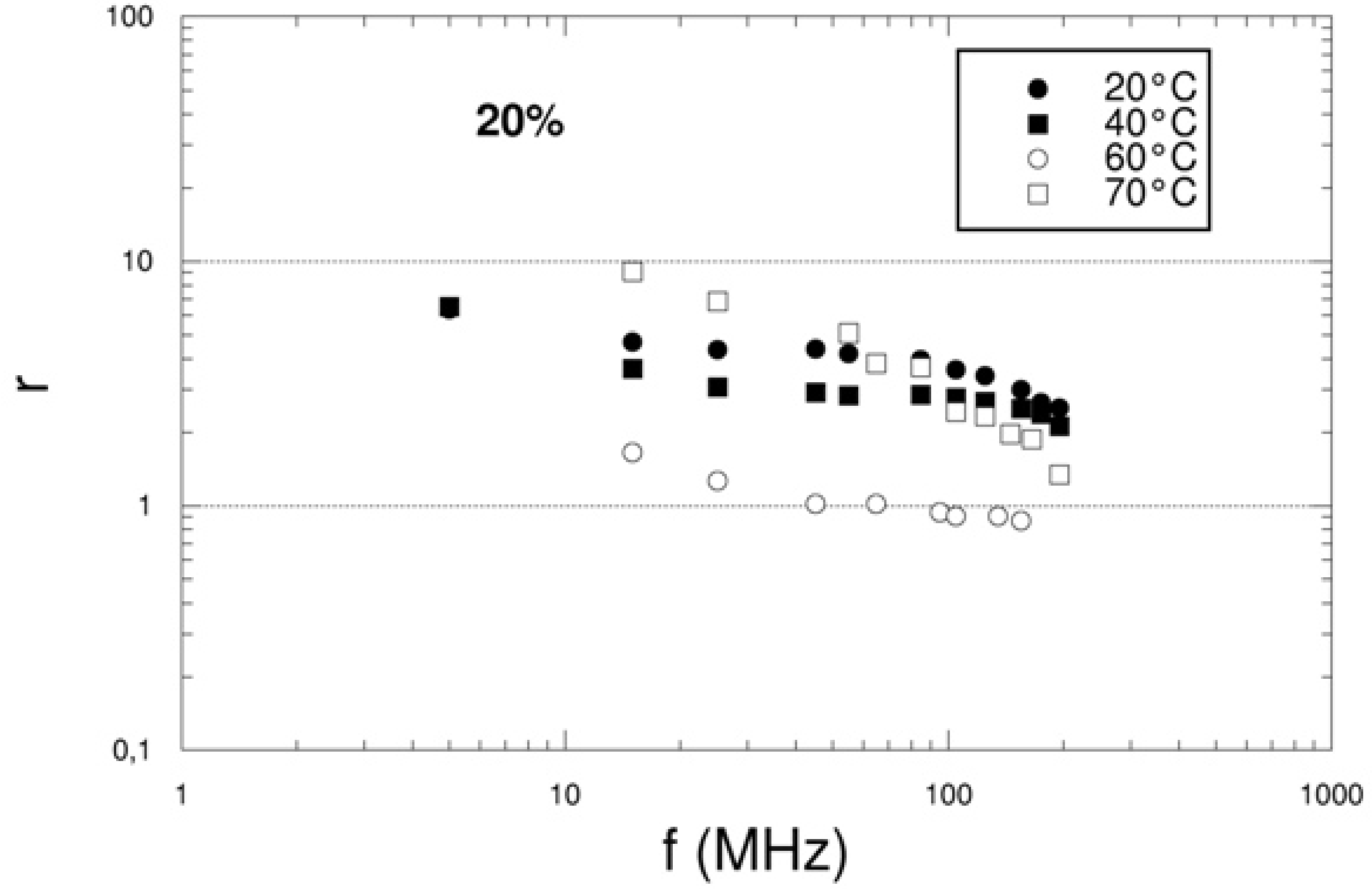

Figure 6b3

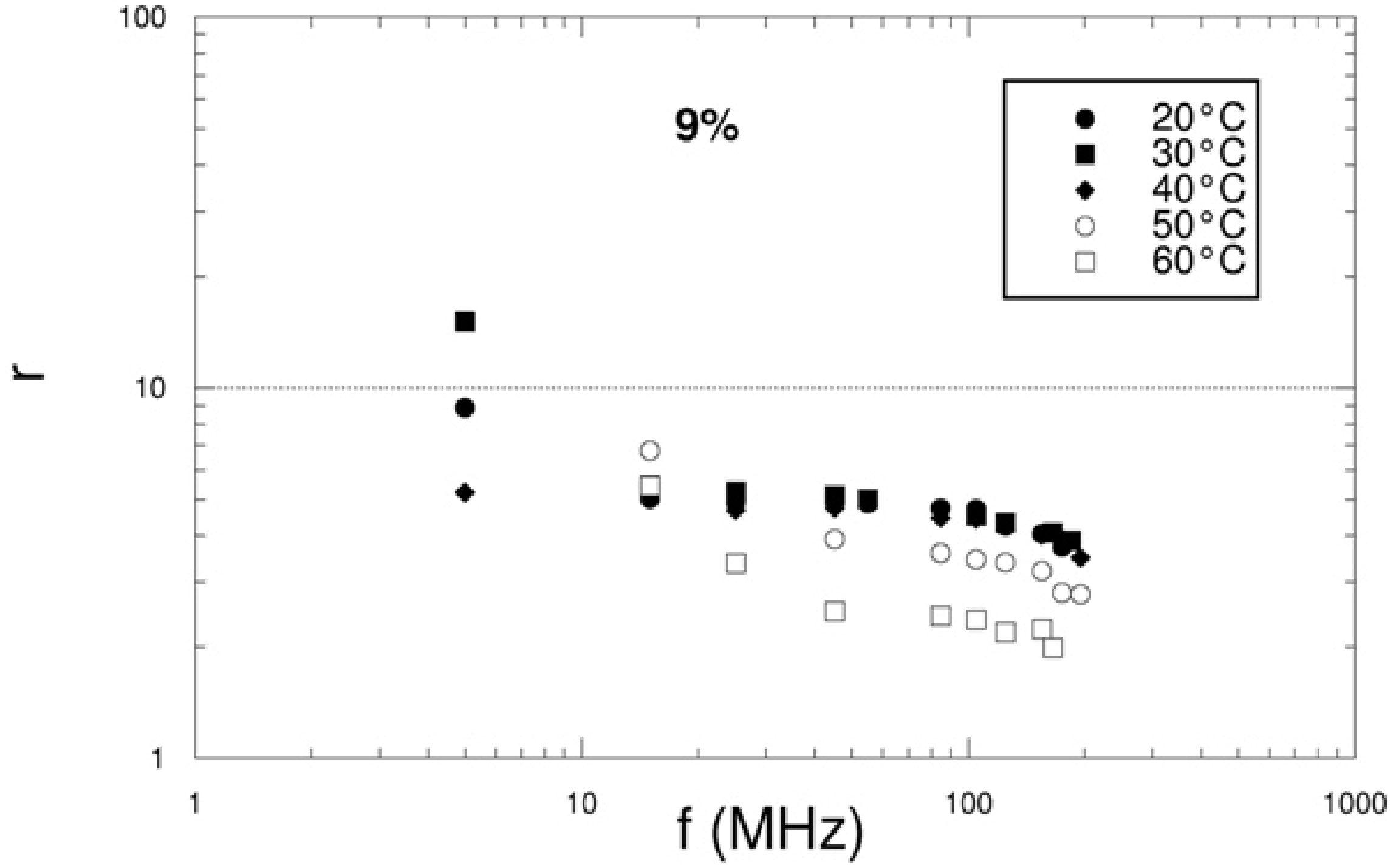

Figure 6b4

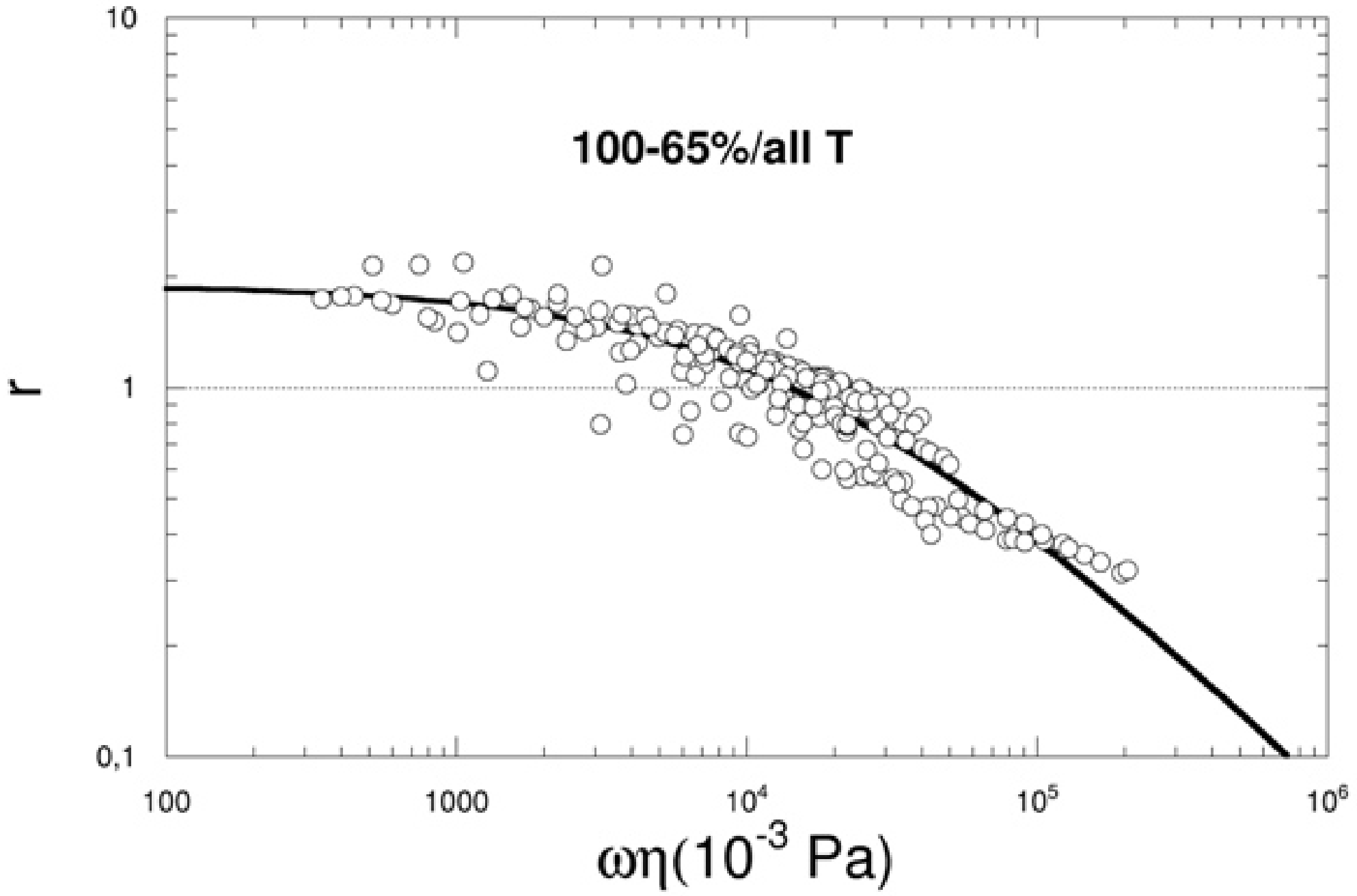

Figure 7

| $\phi$(%) | T (°C) | $B^*_\infty$ | $A^*$ | $G_\infty$(dyn/cm$^2$) | $\beta$ | R[(a)] | $\eta^\circ_v / \eta^\circ_s$ [(b)] | $K_v$(dyn/cm$^2$)[(c)] |
|---|---|---|---|---|---|---|---|---|
| 100-65 | all | 0 | 1.89 | $1.65\cdot 10^{-8}$ | 0.62 | 0.90 | 1.2 | $1.96\cdot 10^{-8}$ |
| 100 | 35-70 | 0 | 1.76 | $3.1\cdot 10^{-8}$ | 0.60 | 0.9682 | 1.01 | $3.08\cdot 10^{-8}$ |
| 90 | 20-70 | 0 | 1.85 | $1.87\cdot 10^{-8}$ | 0.68 | 0.9652 | 1.13 | $2.1\cdot 10^{-8}$ |
| 80 | 20-50 | 0 | 1.72 | $0.51\cdot 10^{-8}$ | 0.79 | 0.9757 | 0.96 | $0.49\cdot 10^{-8}$ |
| | 60-70 | 0 | 2.71 | $0.48\cdot 10^{-8}$ | 0.67 | 0.9815 | 2.27 | $1.1\cdot 10^{-8}$ |
| 65 | 60 | 0 | 2.22 | $1.6\cdot 10^{-8}$ | 0.20 | 0.9969 | 1.63 | $2.6\cdot 10^{-8}$ |
| 40 | 20 | 0 | 1.37 | $1.53\cdot 10^{-8}$ | 0.52 | 0.9907 | 0.50 | $0.76\cdot 10^{-8}$ |
| | 30 | 0.11 | 1.1 | $3.9\cdot 10^{-8}$ | 0.13 | 0.99457 | 0.13 | $0.50\cdot 10^{-8}$ |
| | 40 | 0.24 | 1.99 | $0.3\cdot 10^{-8}$ | 0.48 | 0.99449 | 1.32 | $0.41\cdot 10^{-8}$ |

(a) R is the correlation coefficient of the fitting

(b) $\eta^\circ_v / \eta^\circ_s = 4/3(A^* - 1)$ ; (c) $K_v = (\eta^\circ_v / \eta^\circ_s) G_\infty$

Table 1

[28] F. Sterpone, C. Pierleoni, G. Briganti, M. Marchi, submitted to J. Phys Chem B
[29] C. Sarmoira, D. Blankschtein J. Phys. Chem. (1992) 58, 1978
[30] Eliassi, A.; Modarress, H.; Mansoori, G. A. J. Chem. Eng. Data
1998, 43, 719.
[31] Maisano, G.; Majolino, D.; Migliardo, P.; Venuto, S.; Aliotta, F.;
Magazu`, S. Mol. Phys. 1993, 78, 421.
[32] F. Sterpone, G. Briganti, M. Marchi, C. Pierleoni, Langmuir 20, 4311-4314 (2004)
[33] K. Tasaki, J. Am. Chem. Soc. 118, 8459 (1996)
[34] M.A. Cochran, J.H. Dunbar, A.M. North and R.A. Pethrick, J. Chem. Faraday Trans. II 70 ,215 (1974); V. N. Borodin, S.B. Grigor'ev and I.G. Mikhailov, Sov.Phys. Acoust. 21, 425 (1976).
[35] I. Alig, F. Stieber, A. D. Bakhramov, Yu. S. Manucharov, S. A. Manucharova and V.A. Solovyev, Polymer, 31, 877 (1990); W. Mayer, S. Hoffmann, G. Meier and I. Alig, Phys. Rev. E, 55, 3102 (1997)
[36] G. G. Hammes and T. B. Lewis, J. Phys. Chem. 70, 1610 (1966); L. W. Kessler, W.D. O'Brien Jr and F. Dunn, J. Phys. Chem., 74, 4096 (1970).
[37] O. Metiu and K.F. Freed, J. Chem. Phys., 67, 3303 (1977)
[38] S. Puvvada, D. Blankshtein J. Chem. Phys. 92 (1990) 3710